\documentclass[aps,nofootinbib,showpacs,twocolumn,longbibliography]{revtex4-1}
\usepackage{amstext,amsmath,amssymb,amsfonts,bbm}
\usepackage[latin1]{inputenc}
\usepackage{graphicx}
\usepackage{epstopdf}
\usepackage{amsthm}
\usepackage{tocvsec2}
\usepackage{enumerate}
\usepackage{subfigure}
\usepackage{color}

\usepackage{multirow} \usepackage{rotating}
\usepackage[colorlinks,citecolor=blue,linkcolor=blue,urlcolor=blue]{hyperref}

\def\beq{\begin{equation}}
\def\be{\begin{equation}}
\def\ee{\end{equation}}
\def\bes{\begin{eqnarray}}
\def\ees{\end{eqnarray}}

\begin{document}
\maxtocdepth{subsection}

\title{\Large Divergences and Orientation in Spinfoams}
\author{{\bf Marios Christodoulou, Miklos L{\aa}ngvik, Aldo Riello, Christian R\"oken, Carlo Rovelli\\[-3mm]{\ }}}
\affiliation{Centre de Physique Th\'eorique, F-13288 Marseille, EU}
\pacs{04.60:Pp}

\date{\small\today}

\begin{abstract}\noindent

We suggest that large radiative corrections appearing in the spinfoam framework might be tied to the implicit sum over orientations.  Specifically, we show that in a suitably simplified  context the characteristic ``spike" divergence of the Ponzano-Regge model disappears when restricting the theory to just one of the two orientations appearing in the asymptotic limit of the vertex amplitude.

\end{abstract} 

\maketitle


\section{Introduction} 

Three key open issues in the covariant loop approach to quantum gravity\footnote{See \cite{Rovelli:2011eq} and references therein.} are the presence of two terms of opposite orientation in the asymptotic expansion of the vertex amplitude \cite{Barrett:2009mw}, the role of gauge-invariance \cite{Freidel:2004vi,Baratin:2011tg,Bahr:2009mc}, and the large-spin divergences.\footnote{If the cosmological constant  $\Lambda$ is positive,  the amplitudes of covariant loop quantum gravity are finite \cite{Fairbairn:2010cp,Han:2011vn}. The issue of large-spin divergences for vanishing $\Lambda$ may then seem devoid of physical interest.  It is not so, since, due to the size of $\Lambda$, radiative corrections can still be large,  jeopardizing the regime of validity of the expansion that defines the theory \cite{Rovelli:2011mf}.}  Here we present some indirect evidence that these three issues might be strictly related.  While the relation between divergences and gauge has been pointed out \cite{Freidel:2002dw,Baratin:2011tg}, the connection with orientation is, as far as we know, new.  

In particular, we present some indirect evidence that spinfoam divergences may be generated by the presence of the two terms of opposite orientation in the asymptotic expansion of the vertex amplitude. Pictorially, divergences are generated by geometrical ``spikes" where one of the simplices of the spike has  orientation opposite to the rest of the geometry. Opposite orientation can be related to time (or parity)-inversion \cite{Christodoulou:2012sm} in the Lorentzian theory, hence divergences, or, more precisely, large radiative corrections, might come from large fluctuations of the geometry ``back and forth in time". 

\begin{figure}[h]
\centerline{\includegraphics[height=4cm]{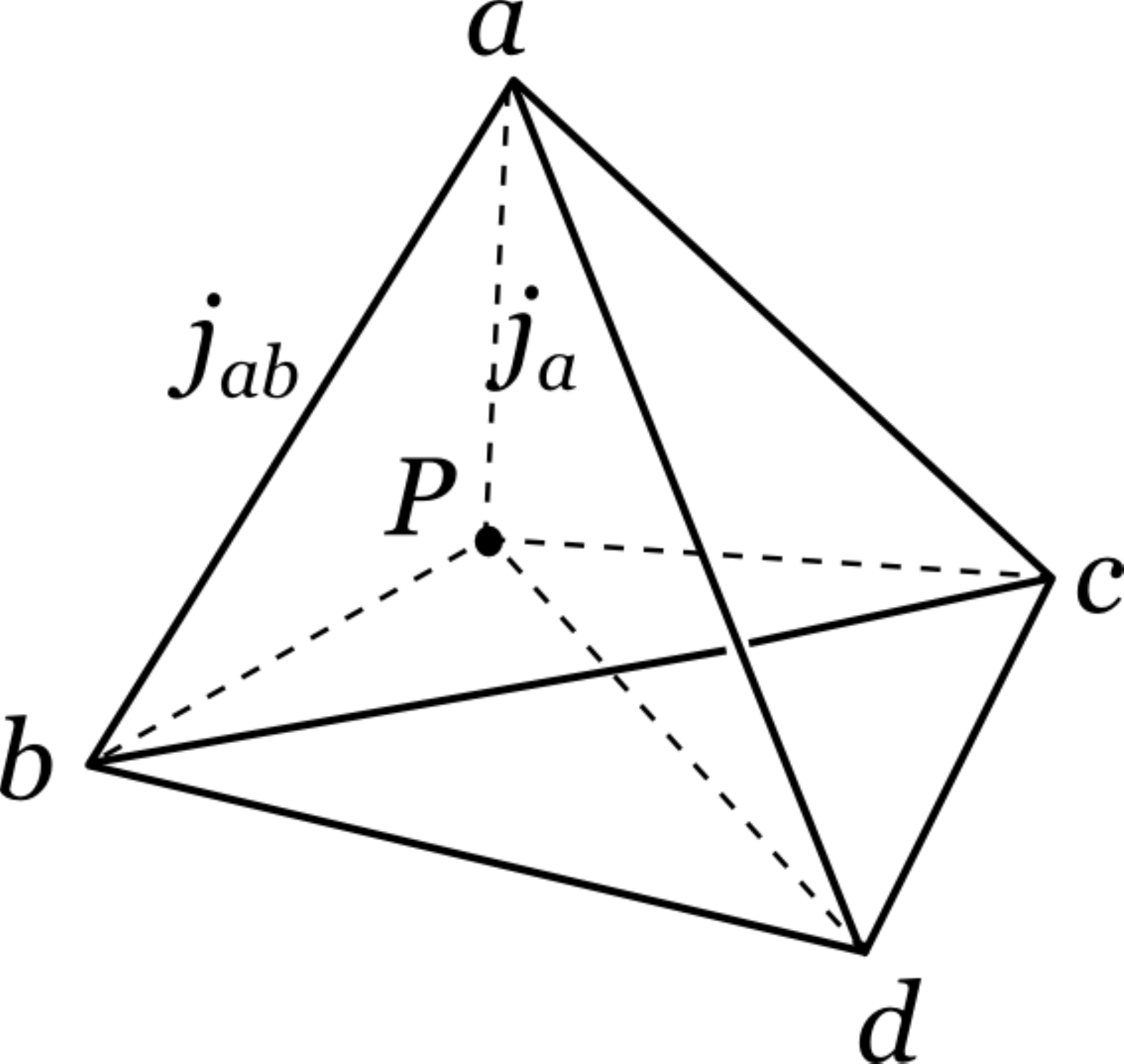}}
\caption{The triangulation $^4 \tau$.} 
\label{F1}
\end{figure}

To illustrate the point, let us start from a puzzle.  Let $^4 \tau$ be the triangulation obtained by splitting a single tetrahedron $\tau$ into four tetrahedra via a 1-4 Pachner move, namely by adding a point $P$ connected to all the vertices of $\tau$, as in Figure \ref{F1}. The Ponzano-Regge amplitude associated to this triangulation is well known to diverge \cite{Ponzano:1968uq}.  This divergence has been associated to the invariance of the amplitude under translations of the point $P$ in the 3d hyperplane defined by the tetrahedron.  Indeed, in the asymptotic expansion of the vertex, the amplitude is determined by the Regge action, and the Regge action is invariant under an infinitesimal 3d translation of the point $P$ \cite{Morse:1991te}.  One is therefore led to identify the divergence with the existence of the gauge orbit generated by moving $P$ in the hyperplane defined by $\tau$.  

However, the Regge action is invariant in this way only as long as the point $P$ is \emph{inside} $\tau$.  Below we explicitly illustrate this point, perhaps under-appreciated, discussing the geometry of Regge calculus.  If the invariance is limited to translations of  $P$ within $\tau$, which is compact, how come the amplitude diverges?

We suggest here that the answer lies in the fact that the asymptotic limit of the Ponzano-Regge amplitude is not the exponential of the Regge action, but rather the sum of two exponentials of the Regge action, taken with \emph{certain} flipped signs. With flipped signs, the invariant contribution comes when $P$ is outside $\tau$.  In other words, the divergence is strictly dependent on the existence of the second term in the expansion of the vertex amplitude. 

The geometrical origin of this second term can be traced to the fact that the asymptotic limit of the Ponzano-Regge model is not truly 3d general relativity in metric variables, but rather 3d general relativity in triad variables, with an action that flips sign under reversal of the orientation of the triad \cite{Rovelli:2012yy}. In three dimensions, it is this action (and not metric general relativity) which is equivalent to $BF$ theory. In turn, $BF$ theory has an additional gauge symmetry with respect to general relativity: the shift $B\to B+d_A\Phi$ (where $A$ is the connection variable: $F=dA+A\wedge A$), which can be shown to be related to the displacement of $P$ all over the hyperplane \cite{Baratin:2011tg}. 

In this paper, we present two arguments that provide some ground for these intuitions. First, we observe that the only classical solutions of Regge theory on the triangulation $^4 \tau$ are those where  $P$ lies inside $\tau$ while in the presence of the flipped-orientation term in the action, there are also classical solutions with $P$ outside $\tau$.   

Second, and more specifically, we study in some detail a simplified form of the asymptotic approximation of the sum that defines the divergence. As will be discussed in more detail, the simplifications consist in trivializing the integration measure and in cutting away a region of the integration domain which is more complicated to control since it does not admit a 4d geometrical embedding. Such simplifications define a quantum Regge calculus for the Regge triangulation $^4 \tau$.  We show that the corresponding amplitude does not diverge if we take only  the Regge term, and diverges if we include the flipped-oriented term.

This is the main technical result of the paper and comes a bit as a surprise, given the intuition that quantum Regge calculus diverges because of the ``spikes".   Here we show that crucial spikes at the root of the divergences are those formed by ``anti-spacetimes", in the language of \cite{Rovelli:2012yy,Christodoulou:2012sm}. These do not exist in conventional Regge calculus. 

The relevance of these observations for the physical four-dimensional quantum gravity is far from granted, but it is suggestive. Looking ahead, we do not think that the moral of our results is that the quantum theory of gravity should be formulated by dropping the flipped term contributions to the dynamics \cite{Engle:2011ps,Engle:2011un,Engle:2012yg,Rovelli:2012yy}. Rather, it should be better understood what kind of relation exists among ``flipped-orientation" (or ``anti-spacetime") quantum fluctuations of the geometry, gauge-invariance, and radiative corrections.

The paper is organized as follows.  We start by illustrating the geometry of the spikes in Regge calculus in 2d and in 3d in Section \ref{secII}.  In Section \ref{secIII}, we discuss the Ponzano-Regge model and its large spin limit, and in Section \ref{secsadd}  we show how the flipped-orientation terms in the action give rise to new classical solutions, with $P$ outside $\tau$. In Section \ref{bound}, we define the (simplified) integrals corresponding to the amplitudes in a quantum Regge calculus with and without the flipped terms and show that the second amplitude is bounded, while the second is not. Finally we conclude in Section \ref{secconc}. 

\section{Spikes in conventional Regge calculus}\label{secII} 

We start with an illustration of the geometry of the spikes in 2d and 3d Regge calculus, as this is sometimes a source of confusion. We will deal with spikes which can be embedded in one more dimension (i.e. in 3d and 4d respectively). Remark that this is not the most general possible spike (see Section \ref{bound} for more details).

\subsection{Two dimensions} 

\begin{figure}[h]
\centerline{\includegraphics[scale=.23]{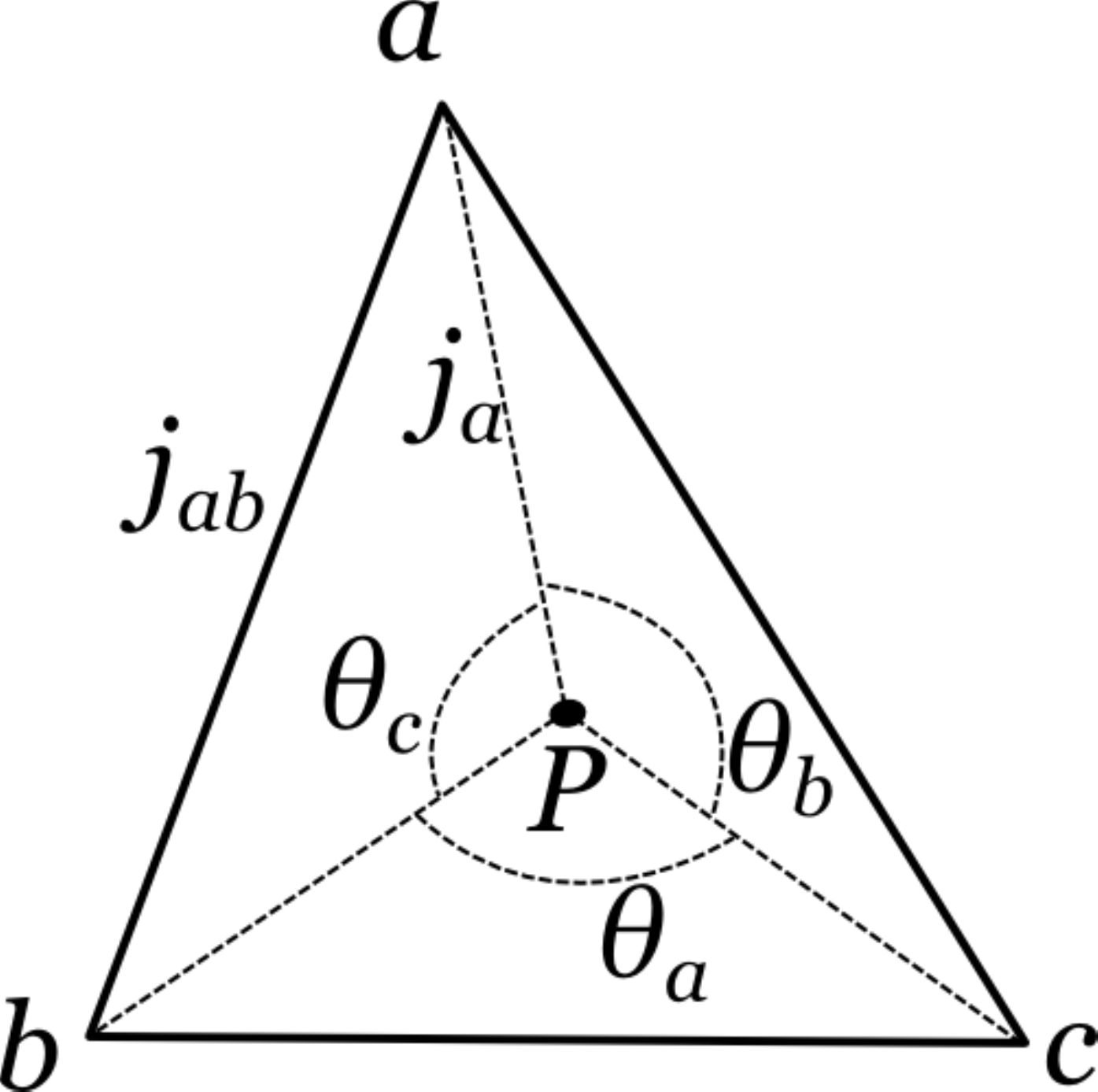}\hspace{2em} \includegraphics[scale=0.23]{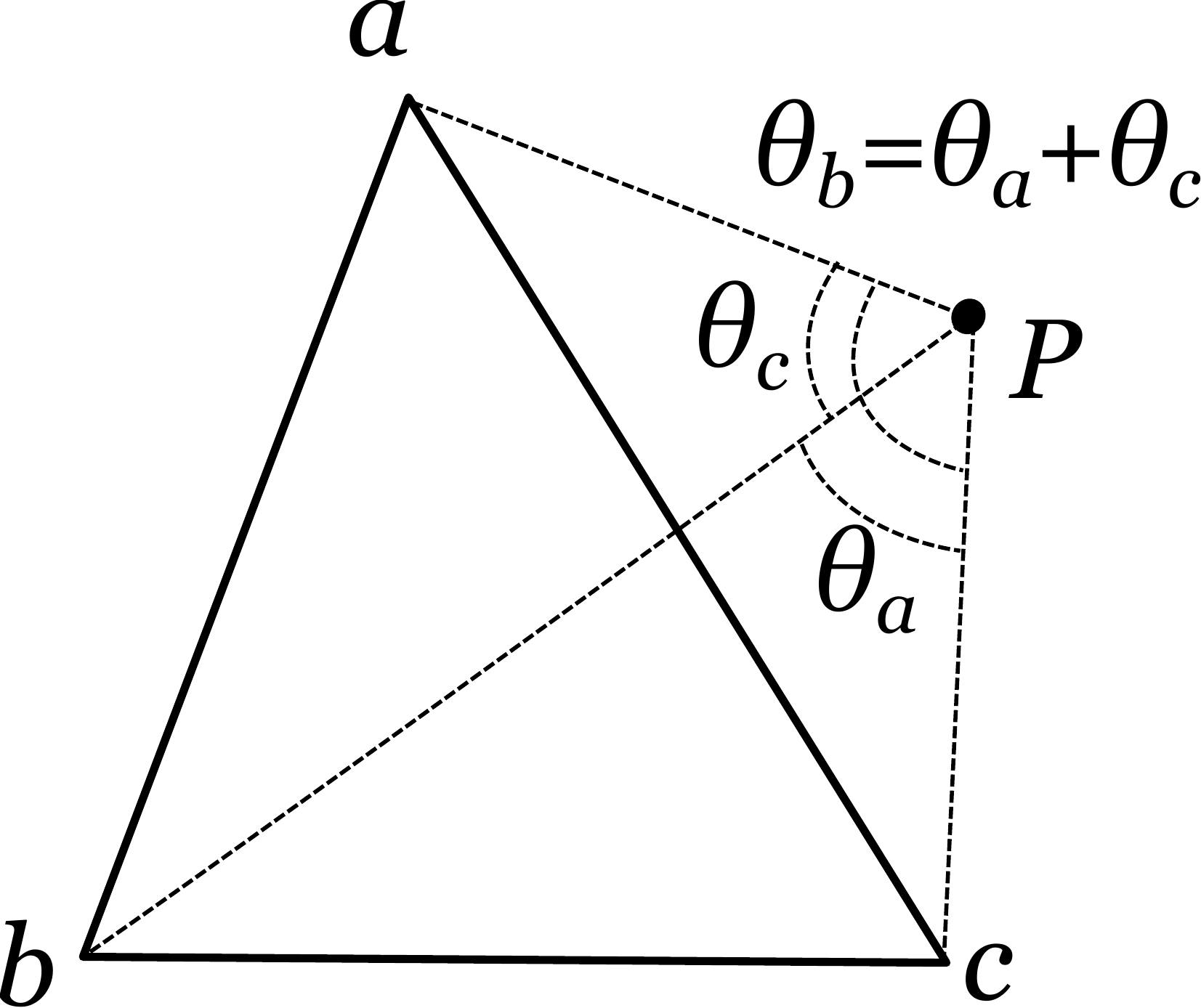}}
\caption{2d triangulation.  Left: $P$ inside $\tau$ and no curvature at $P$. Right: $P$ outside the plane of $\tau$ and curvature at $P$.}
\label{tri}
\end{figure}

Consider the 2d plane $z=0$ immersed in 3d Euclidean space with coordinates $(x,y,z)$. Consider a triangulation of this plane, containing a triangle $\tau$, whose sides have lengths $j_{ab},\; a,b=1,2,3$. Pick a point $P$ in $\mathbb{R}^3$ at a distance $j_a$ from the vertices of $\tau$. By joining $P$ with the three vertices of  $\tau$, we obtain three triangles $\tau_a$, each determined by $P$ and one of the three sides of $\tau$.  The 1-3 Pachner move $\tau\to {}^3\tau=\{\tau_a\}$ modifies the original (flat) triangulation into a triangulation which is in general curved. In particular, there is discrete (distributional) curvature at $P$, given by the deficit angle
\be
  \delta_P:=2\pi-\sum_a\theta_a,
  \ee
where $\theta_a$ is the angle at $P$ of the triangle $\tau_a$.\footnote{\label{notation}Switching for the moment to a more general notation: the angle $\varphi_{pq}$ between the two sides $p$ and $q$ of a triangle with sides $p,q,r$ is given by 
\be
\cos{\varphi_{pq}}:={\frac{p^2+q^2-r^2}{2pq}} \quad \text{and} \quad \varphi_{pq}\in[0,\pi].
\label{angolo}
\ee
}
The discrete curvature $\delta_P$ is a well-defined continuous function of the position of $P$ in $\mathbb{R}^3$. It vanishes if $P$ lies in $\tau$ and increases monotonically if $P$ moves perpendicularly away from $\tau$.  Therefore if $P$ lies in $\tau$ and we consider infinitesimal shifts in its position, there are two directions that do not change the curvature (see the left side of Figure \ref{tri}) and one that does.

Now, what is the value of the curvature if $P$ lies on the plane $z=0$, but outside $\tau$?  A moment of reflection shows that in this case the curvature does \emph{not} vanish.  This can be seen for instance rather simply using continuity: consider a situation where the three lengths $j_a$ are large compared to the size of $\tau$. Then each of the three angles $\theta_a$ is small, and the deficit angle is near $2\pi$. If we move  $P$ continuously toward the plane $z=0$ (but still far from $\tau$), the three angles remain small and $\delta_P$ remains near  $2\pi$. On $z=0$, one of the three angles, say $\theta_b$, will become equal to the sum of the other two (see right side of Figure \ref{tri}), but this does not change the fact that the curvature at $P$ is high.  

It is important to observe that the topology of the triangulation is not affected by this degenerate case.  The degeneracy is in the embedding of the 2d triangulated surface into $\mathbb{R}^3$, not in the topology of the triangulation, which is determined by its intrinsic geometry and which remains $\mathbb{R}^2$.  Confusing the intrinsic topology of the 2d triangulation with that induced by its embedding in $\mathbb{R}^3$ is the source of the misleading intuition that the triangulation with $P$ in the plane but outside $\tau$ is flat. 

The only part of the plane where the curvature is constant is the interior of $\tau$. 

\subsection{Three dimensions} 

\begin{figure}[h]
\centerline{\includegraphics[height=4.5cm]{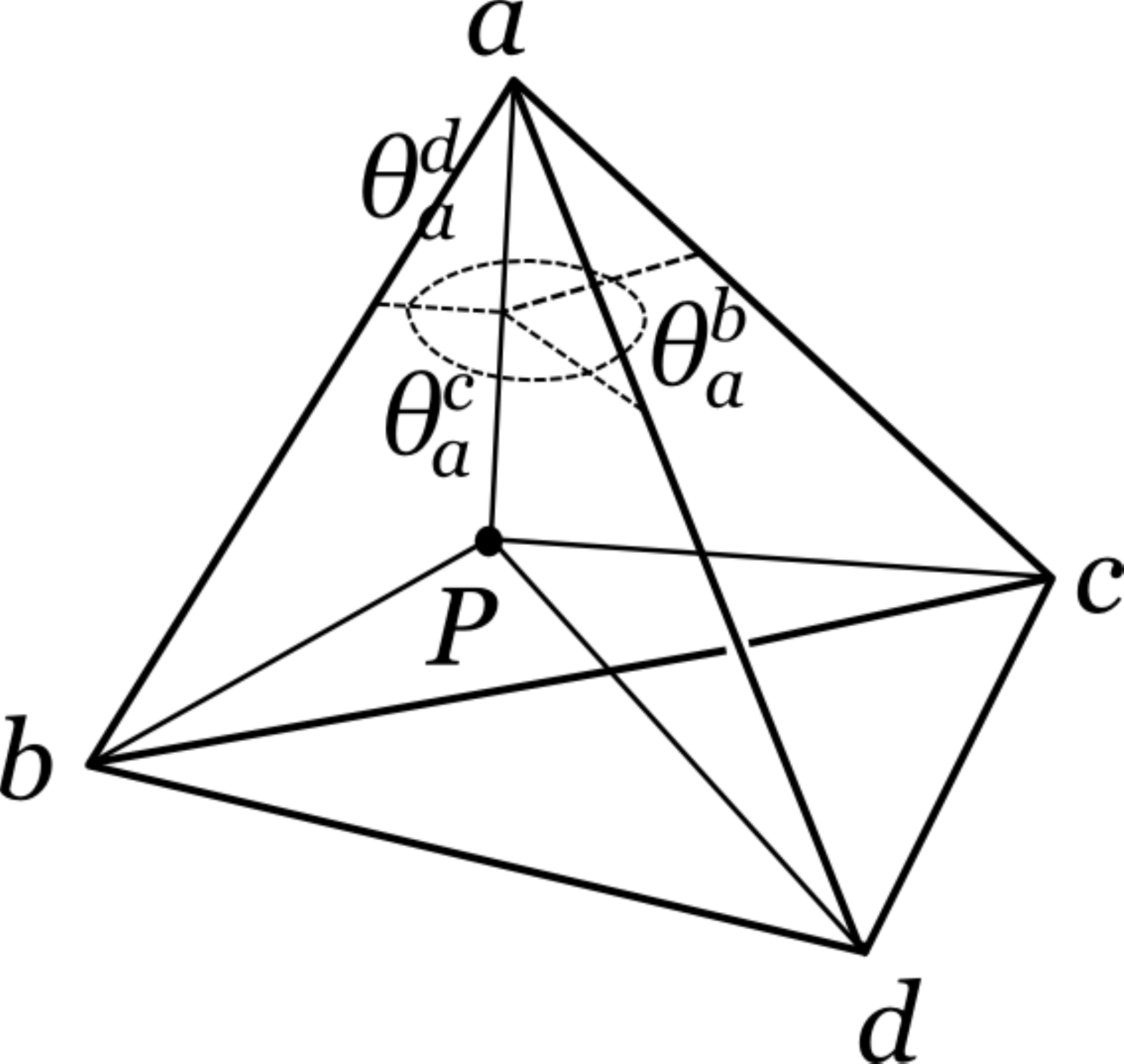}}
\caption{Absence of curvature in 3d triangulation with $P$ inside $\tau$.} 
\label{f3}
\end{figure}

Let us repeat now the same construction in 3d.  Consider a tetrahedron $\tau$ lying in the 3d hyperplane $s=0$ immersed in 4d Euclidean space with coordinates $(x,y,z,s)$, with sides having lengths $j_{ab},\; a,b=1,2,3,4$. Pick a point $P$ in $\mathbb{R}^4$ and let  $j_a$ be the distances of $P$ from the vertices of $\tau$. By joining $P$ with the four vertices of  $\tau$, we obtain four tetrahedra $\tau_a$, each determined by $P$ and one of the four triangles which bound $\tau$.  The 1-4 Pachner move $\tau\to{}^4\tau=\{\tau_a\}$ splits the original tetrahedron into a triangulation (Figure \ref{F1}) which is generally curved. In particular, there is distributional curvature $\theta_a$ at each of the four segments joining $P$ with the vertices of $\tau$, given by the deficit angles
\be
  \delta_a:=2\pi-\sum_{b\ne a}\theta_a^b,
\ee
where $\theta_a^b$ is the internal dihedral angle at the edge $a$ of the tetrahedron $\tau_b$ (Figure \ref{f3}).\footnote{Switching to the notation of footnote \ref{notation}: the internal dihedral angle at the side $a$ of a tetrahedron with three sides $a,b,c$ joining in a vertex is given by 
\be
 \cos{\theta_{a}}:={\frac{\cos{\varphi_{bc}}-\cos{\varphi_{ab}}\cos{\varphi_{ac}}}{\sin{\varphi_{ab}}\sin{\varphi_{ac}}}},
\ee
where $\varphi_{ab}\in[0,\pi]$ is the angle formed by the segments $a$ and $b$, given in terms of the sides in \eqref{angolo}.}

\begin{figure}[h]
\centerline{\includegraphics[height=4.5cm]{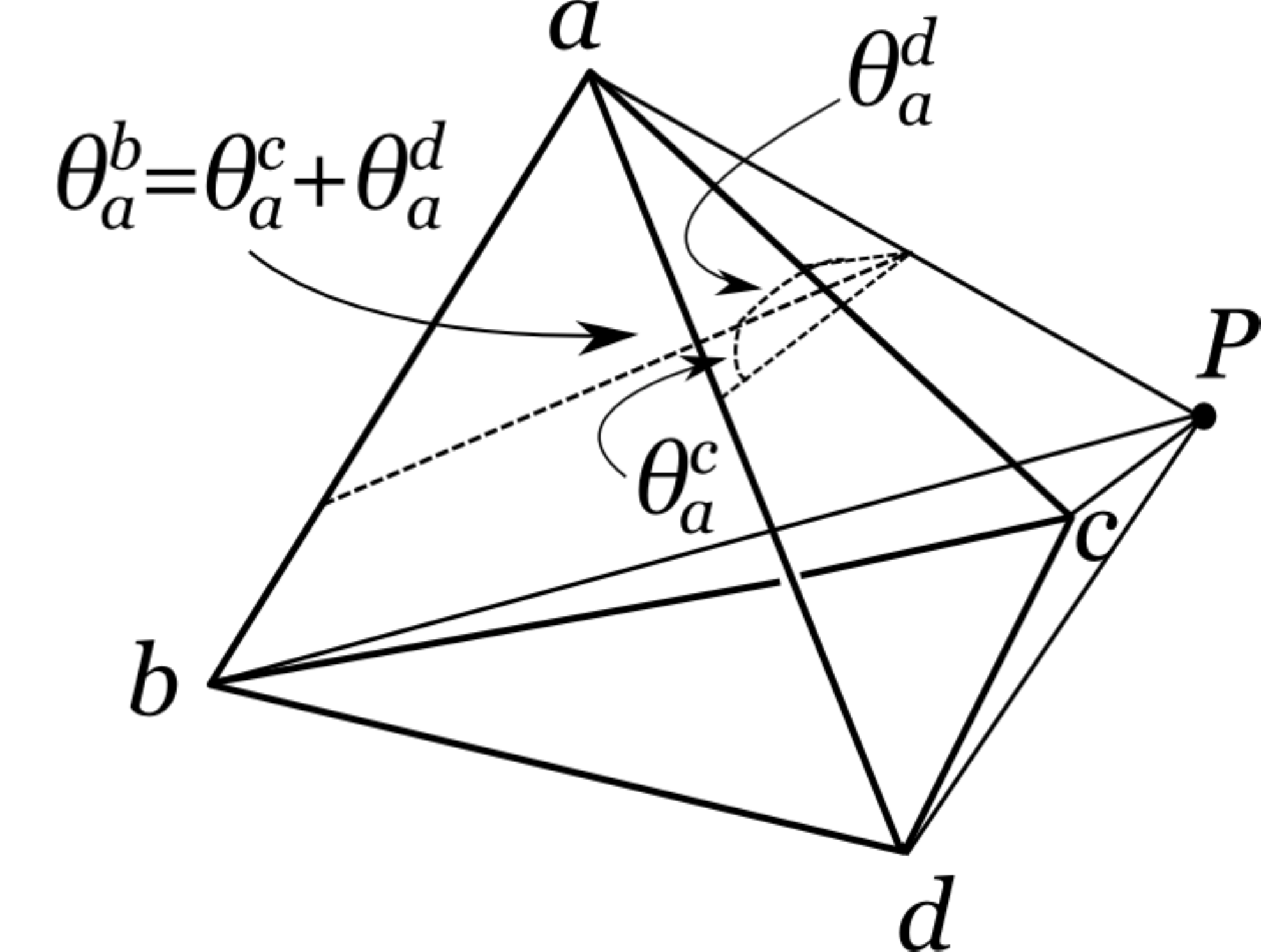}}
\caption{Curvature in 3d triangulation with $P$ outside $\tau$.} 
\label{ch1}
\end{figure}

Again, the deficit angle is a well-defined continuous function of the position of $P$ in $\mathbb{R}^4$. It  vanishes if $P$ lies in $\tau$ (Figure \ref{f3}) and becomes non zero when $P$ is taken perpendicularly away 
from $\tau$ in $\mathbb{R}^4$ (Figure \ref{ch1}).  Therefore if $P$ lies in $\tau$ and we consider infinitesimal shifts in its position, there are three directions that do not change the curvature, and one that does.

Now, what is the value of the curvature if $P$ lies in the hyperplane $s=0$, but outside $\tau$?  As before, in this case the curvature does \emph{not} vanish.  As in the two dimensional case, this can be shown by continuity. However the argument is more subtle, since the dihedral angles do not necessarily go to zero as $P$ goes to infinity; nevertheless, as it will be shown in section \ref{bound}, in this limit the sum of the three dihedral angles associated to the edges $(bP),(cP)$ and $(dP)$ of the tetrahedron $\tau_a$ goes to $\pi$; thus, the sum of all twelve such dihedral angles is $4\pi$, and not $4\times2\pi$ (a full turn for each internal edge of the triangulation) as one would expect for the flat case. Therefore, this configuration represents necessarily a curved Regge geometry. If we now move continuously $P$ into the plane $s=0$, the previous argument continues to hold. When $s=0$, one of the three dihedral angles becomes equal to the sum of the other two dihedral angles relative to the same edge (see Figure \ref{ch1}); nevertheless, this does not change the fact that the curvature is non-zero.

As before, the topology of the triangulation is not affected by this degenerate case: the degeneracy is in the embedding of the 3d triangulated surface into $\mathbb{R}^4$, not in the topology of the triangulation, which remains $\mathbb{R}^3$.\\

The only part of the hyperplane where the curvature remains constant and vanishes is the interior of $\tau$. 

\section{The Ponzano-Regge amplitude in the spin basis}\label{secIII}

Let us now come to the Ponzano-Regge amplitude. For a given triangulation formed by tetrahedra $v$, triangles $e$ and edges $f$,\footnote{The conventional ($v,e,f$)  notation is standard and derives from the dual or ``spinfoam" representation of the triangulation formed by vertices, edges and faces.} this is given in the spin basis by \cite{Ponzano:1968uq,Barrett:2008wh}
\begin{align}
  W(j_l):=\sum_{j_f}{}'\prod_f & (-1)^{2j_f}(2j_f+1)  \times\notag\\
&\times\prod_{e} (-1)^{j_{e_1}+j_{e_2}+j_{e_3}}\prod_v \{ 6j \}_v. \label{PR}
\end{align}
Here $j_f\,$ are the spins associated to (any of) the edges of a given triangulation, $j_l$ are those associated uniquely to its boundary edges, $\{ 6j \}_v$ is the Wigner 6-$j$ symbol associated to the tetrahedron $v$, and $e_i$ are the three sides of the triangle $e$. The primed sum runs over the bulk spins alone.

The sign factors are necessary for the topological invariance of the theory as well as for the recovery of general relativity in the continuous limit (see  \cite{Barrett:2008wh}), and will play a key role below.

For the triangulation $^4 \tau$, the Ponzano-Regge amplitude reads:  
\be
  W(j_{ab})=N(j_{ab})\sum_{j_a} \prod_a (-1)^{j_a}(2j_a+1) \prod_a  \left\{ 
\begin{tabular}{@{} ccc @{}}
$j_b$ & $j_c$ & $j_d$ \\ 
$j_{cd}$ & $j_{db}$ & $j_{bc}$
\end{tabular}
\right \},
  \label{esto}
\ee
where $N(j_{ab}):=\prod_{(ab)} (-1)^{j_{ab}}(2j_{ab}+1)$ depends only on the boundary edges. In writing this expression we used the fact that $2j_a\in\mathbb{N}$, as well as the convention according to which different indices take different values (convention that will be used throughout the whole article). It is well known since the work of Ponzano and Regge that this sum diverges.  This can be seen by Fourier transforming from the spins to $SU(2)$ group variables  (see \cite{Rovelli:2011eq}), which gives 
\be
  W \sim \int_{SU(2)^4} dh_{ab} \prod_a \delta_{SU(2)}(h_{bc}h_{cd}h_{db})\sim \delta_{SU(2)}(1).
\ee

This divergence goes like $j^3$, suggesting that it is related to a 3d volume divergence. It is tempting to relate it to the invariance of the Regge amplitude under shifts of the point $P$ in the 3d hyperplane of the tetrahedron $\tau$.  Indeed, the divergence comes obviously from the large $j_a$'s region, where we can use the asymptotic value for the Wigner 6-$j$ symbols:\footnote{Rigorously speaking, this asymptotic expression is available only if \emph{all} the $j$'s go to infinity \emph{uniformly}, the boundary spins $j_{ab}$ included. }

\be
 \{ 6j \} \sim \frac{1}{\sqrt{3\pi V}}\cos\left(\sum_{f}\Theta_{f} j_{f}+\frac\pi4 \right),
\ee
where $\Theta_{f}$ is the external dihedral angle at the edge $f$ of a tetrahedron having sides with lengths $j_{f}$, and $V$ is the volume of this tetrahedron. Remarkably, if the six $j_f$'s are such that they do not close geometrically into triangles, then the corresponding 6-$j$ symbol is ill-posed and set to zero; moreover, if they do geometrically close into triangles, but they do not into a tetrahedron, then the corresponding 6-$j$ symbol in the large $j$ limit is exponentially suppressed.

Inspired by this asymptotic behaviour, in a notation suitable for $^4\tau$, we define
\be
A_{\pm}^a = \frac{1}{\sqrt{12\pi V_a}} \exp\left[{\pm i\left(\sum_{\varepsilon\subset\tau_a}\Theta_{\varepsilon}^a j_{\varepsilon}+\frac\pi4\right)}\right],
\ee
where $V_a$ is the volume of the tetrahedron $\tau_a$, and where the sum runs over its edges.\footnote{For example the edges of $\tau_4$ are $\{(1P),(2P),(3P),(12),$ $(23),(31)\}$, and their lengths are $\{j_1,j_2,j_3,j_{12},j_{23},j_{31} \}$, respectively.}
Then, the asymptotic limit of \eqref{esto} can be split into sixteen terms
\begin{eqnarray}
  W(j_{ab})&\sim&N(j_{ab})\sum_{j_a}  \prod_a (-1)^{j_a}(2j_a+1) \prod_a (A^a_+ + A^a_-)\nonumber \\
  &\sim&\ \ \sum_{\pm\pm\pm\pm} W_{\pm\pm\pm\pm}(j_{ab}).
  \label{estospli}
\end{eqnarray}
The analysis of these terms will be the central topic of the next section.

\section{The $W_{\pm\pm\pm\pm}$'s and their saddle points}\label{secsadd}

In this section we give an expression of the $W_{\pm\pm\pm\pm}$ for each of the four relevant combinations of $\pm$ signs, we discuss the continuum (large $j$) limit and the relation of the latter to the classical equation of motion.

Using the definition given implicitly through equation \eqref{estospli} and by writing the $(-1)^j$ factors as $e^{ik\pi j}$, $k\in(2\mathbb{Z}+1)$, (more will be said about this ambiguity later) one finds
\begin{align}
W_{\pm\pm\pm\pm}&=N'(j_{ab})\sum_{j_a}\mu[j_f]e^{i\sum_{(ab)} j_{ab}\left[\pm_c\Theta_{ab}^c\pm_d\Theta_{ab}^d +k\pi  \right]  }\times\notag\\
&\qquad\qquad\qquad\times e^{i \sum_a j_a \left[ \pm_b \Theta^b_a \pm_c \Theta^c_a \pm_d \Theta^d_a  +k\pi\right]}\notag\\
&=:N'(j_{ab})\sum_{j_a}\mu[j_f]e^{iS_{\pm\pm\pm\pm}}\;,\label{Spm}
\end{align}
where $N'(j_{ab}):=\prod_{(ab)}(2j_{ab}+1)$, $\mu[j_f]:=\prod_a\frac{2j_a+1}{\sqrt{12\pi V_a}} $, and $\pm_a$ is plus or minus according to which one between $A_+^a$ and $A_-^a$ appears in $W_{\pm\pm\pm\pm}$. Finally in \eqref{Spm} we defined an ``action" $S_{\pm\pm\pm\pm}$ for each $W_{\pm\pm\pm\pm}$. We can now switch from external ($\Theta$) to internal dihedral angles ($\theta$) by using the relation: $\Theta=\pi-\theta$. This straightforwardly leads to:
\begin{align}
S_{++++}=&\sum_{(ab)} j_{ab}\left[(2+k)\pi-\theta_{ab}^c-\theta_{ab}^d \right] +\notag\\
&+ \sum_a j_a \left[ (3+k)\pi -( \theta^b_a + \theta^c_a + \theta^d_a )\right],\label{S+}\\
\notag\\
S_{+++-}=&\dots
+\sum_{a\neq4} j_a \left[ (1+k)\pi -( \theta^b_a + \theta^c_a - \theta^4_a )\right]+\notag\\
&+ j_4 \left[ (3+k)\pi -( \theta^1_4 + \theta^2_4 + \theta^3_4 )\right],\label{S+++-}\\
\notag\\
S_{++--}=& \dots +\sum_{a=1,2} j_a \left[(-1+k)\pi -( \theta^b_a - \theta^3_a -\theta^4_a )\right]+\notag\\
&+\sum_{c=3,4} j_c \left[ (1+k)\pi -( \theta^1_c + \theta^2_c - \theta^d_c )\right],\label{S++--}\\
\notag\\
S_{+---}=& \dots+j_1 \left[ (-3+k)\pi +\theta^2_1 + \theta^3_1 + \theta^4_1 \right]+\notag\\
& + \sum_{a\neq1} j_a \left[ (-1+k)\pi -( \theta^1_a - \theta^b_a - \theta^c_a )\right],\label{S+---}\\
\notag\\
S_{----}=&\sum_{(ab)} j_{ab}\left[(-2+k)\pi+\theta_{ab}^c+\theta_{ab}^d \right] +\notag\\
&+ \sum_a j_a \left[ (-3+k)\pi + \theta^b_a + \theta^c_a + \theta^d_a \right].\label{S-}
\end{align}
In (\ref{S+++-}-\ref{S+---}) dots indicate that we omitted the terms associated to the boundary spins, which are however irrelevant for the following.

Now, we come to the specific choice of $k\in(2\mathbb{Z}+1)$. This choice is imposed by the fact that we want to recover the Regge action in the large-$j$ continuum limit. We see that only \eqref{S+} and \eqref{S-} are good (mutually exclusive) candidates for this. In fact they are the only actions which contain for every internal edge the sum of all the three associated dihedral angles which appears in the definition of the deficit angle. Therefore, if we want to make $S_+:=S_{++++}$ essentially equal to the Regge action $S_R=\sum_f j_f\delta_f$ we have to choose $k=-1$; conversely if we want to make $S_-:=S_{----}$ essentially equal to (minus) the Regge action, we have to choose $k=1$.\footnote{In fact this is a simplified version of a more general argument concerning the Ponzano-Regge amplitudes \eqref{PR} for an arbitrary triangulation. It is straight forward to see that in \eqref{PR} the following substitution must be done in order to get the Regge action in the asymptotic (continuum) limit: $(-1)^{2j_f}\to e^{-i2k\pi j_f}$ and $(-1)^{j_{e_1}+j_{e_2}+j_{e_3}}\to e^{ik\pi (j_{e_1}+j_{e_2}+j_{e_3})}$, with $k=\mp 1$ according to which one between $W_+$ or $W_-$ should lead to the Regge action in the limit ($W_\pm$ being the natural generalization to an arbitrary triangulation of $W_{++++}$ and $W_{----}$ defined in the context of $^4\tau$).  } 

The two choices are symmetric, i.e. relate $S_{\pm\pm\pm\pm}$ in one case to $-S_{\mp\mp\mp\mp}$ in the other. Choose once and for all 
\be
k=-1, \text{ so that } S_+\sim S_R.
\ee Notice that once this choice is done, no symmetry is any more present among the previous actions (\ref{S+}-\ref{S-}). This fact entails an important consequence, which is important for the coherence of our results in section \ref{bound}: i.e. \emph{only the actions $S_+$ and $S_{+++-}$ have a (large set) of classical solutions}. In the following of this section we shall justify and discuss this statement.\\

As it is customary, we define the classical solutions to be described by the stationary points of the action, picked out by letting $j_a$'s go to infinity.\footnote{This is in fact equivalent to letting $\hbar$ going to zero.} Hence, the classical equations of motion for (\ref{S+}-\ref{S-}) are respectively:
\begin{align}
S_{+}\sim S_R\; &: \; \delta_a:=2\pi-(\theta_a^b+\theta_a^c+\theta_a^d)=0;\label{sadd+}\\\notag\\
S_{+++-}\; &:\; \left\{\begin{array}{ll} \theta_a^b+\theta_a^c=\theta_a^4, & a\neq4 \\ \delta_4=0\end{array}\right.;\label{sadd+++-}\\\notag\\
S_{++--}\;&:\;\left\{\begin{array}{ll} \theta_a^3+\theta_a^4=\theta_a^b+2\pi, & a=1,2 \\ \theta^1_c+\theta^2_c=\theta^d_c& c=3,4\end{array}\right.;\label{sadd++--}\\\notag\\
S_{+---}\;&:\;\left\{\begin{array}{ll} \theta_1^2+\theta_1^3+\theta_1^4=4\pi\\ \theta^b_a+\theta^c_a=\theta^1_a+2\pi\end{array}\right.;\label{sadd+---}\\\notag\\
S_{-}\; &: \; \theta_a^b+\theta_a^c+\theta_a^d=4\pi.\label{sadd-}
\end{align}
As it is well known, Equation \eqref{sadd+} is the flatness condition expressed in the language of Regge geometry, i.e. by the vanishing of all deficit angles; solutions to these equations are the quadruplets of lengths $j_a$ given by the distances between the vertices of $\tau$ and a point $P$ situated anywhere \emph{inside} $\tau$ (see Section \ref{secII}).

Solutions to Equation \eqref{sadd+++-} are similarly parametrized by the position of the point $P$ in a specific region of $\mathbb{R}^3$, via an embedding of the triangulation. In this case, however, such a region, say $\mathcal R_4$, is non-compact and lies outside the tetrahedron $\tau$. $\mathcal R_4$ is the disconnected union of two regions, $\mathcal{R}_4^f$ and $\mathcal{R}_4^v$; loosely speaking $\mathcal R_4^f$ correspond to the sector of $\mathbb{R}^3$ from which one cannot ``see" the fourth vertex of $\tau$, since it is  ``hidden" by the tetrahedron $\tau$ itself (see Figure \ref{ch1}), while $\mathcal R_4^v$ is the point reflected of $\tau\cup\mathcal R_4^f$ with respect to the fourth vertex of $\tau$.

The saddle point equations \eqref{sadd++--} for $S_{++--}$ are slightly more involved. First, recall that $0\leq\theta_a^b\leq\pi$, then by the first pair of equations one gets $\theta_2^c=\theta_1^c=\pi$; this is possible only if $P$ lies on the face $(12d)$, with $d\notin\{1,2,c\}$ of $\tau$; since $d\in\{3,4\}$, this means that $P$ lies on the edge $(12)$; the second pair of equations in \eqref{sadd++--} has the same solution, in fact it is satisfied only if $P$ lies at the same time in $\mathcal{R}_3$ and $\mathcal{R}_4$, i.e. if $P$ lies on the edge $(12)$. Thus, the classical solutions to the action $S_{++--}$ are parametrized by $P\in(12)$. 

Finally, Equations \eqref{sadd+---} and \eqref{sadd-} have no classical solution. To see this, just recall once more that  $0\leq\theta_a^b\leq\pi$, and remark that therefore $\theta_a^b+\theta_a^c+\theta_a^d\leq3\pi<4\pi$.\\

Thus, we showed that the classical solutions to the asymptotic $j\gg1$ Ponzano-Regge model come mainly from two different sectors of the asymptotic limit (in fact the solutions stemming from $S_{++--}$ are given by a compact region of co-dimension three with respect to the phase space $\{j_a\}$; therefore, they are practically irrelevant for any concrete purpose). Moreover the solutions stemming from the $S_+\sim S_R$ sector are all gauge equivalent since they describe the very same geometry. However, the corresponding gauge orbit is compact, and for this reason it is hard to explain the model's divergences by saying that they arise from the integration over such gauge equivalent solutions. On the other hand, there exist four infinite non-compact sets of classical solutions which have co-dimension one, stemming from the terms of the type $S_{+++-}$. It is then quite tempting to associate the divergence of the Ponzano-Regge amplitude of $^4\tau$ to the existence of this three dimensional hyper-surfaces of saddle points (recall that $W[^4\tau]$ diverges like $j^3$). \\

In the following section, we explore this hypothesis from quite a different point of view, by studying explicitly (even if under some simplifying assumptions) a certain continuum version of the $W_{\pm\pm\pm\pm}$.

 \section{Boundedness of the Regge spike}\label{bound}

We now come to the main technical result of this paper.  We study a continuum and simplified version of the amplitude $W_+$ and we show that it is bounded. The argument then fails for the corresponding integral in the $W_{+++-}$ case (but stays valid in the remaining cases).  

We first rewrite $W_+$ in the limit where the $j$'s become continuum variables, dubbed $l_a$ and $l_{ab}$, in the case of bulk and boundary edge lengths respectively:

\begin{align}
W_+(l_{ab})\approx N'(l_{ab}) \int_\Omega d\mu[l_{a}]& e^{i\sum_{(ab)}l_{ab}\left[\pi-\left(\theta_{ab}^c+\theta_{ab}^d\right)\right]}\times\notag\\
&\times e^{i\sum_a l_a\left[2\pi-\left(\theta_a^b+\theta_a^c+\theta_a^d\right)\right]},\label{cont}
\end{align}
where the domain of integration $\Omega$ is given by the quadruplets of lengths that satisfy two sets of conditions, which assure from the algebraic point of view that the 6-$j$ symbols are well defined and from the geometrical point of view that the corresponding segments close into triangles and tetrahedra following the combinatorics of the triangulation $^4\tau$.\footnote{Such conditions are mathematically expressed \cite{CM} via the positivity of the Cayley-Menger determinant for the 2- and 3-simplices. The Cayley-Menger determinant evaluated on the edges of a geometrical $n$-simplex equals the square of its volume. For a 2-simplex, i.e. triangles, the above mentioned condition is equivalent to the triangular inequalities.\label{piede}} Remark that no condition is required for \emph{all} of them to close at the same time, neither in $\mathbb{R}^3$ nor in $\mathbb{R}^4$. 

The simplification we consider is first to replace the non-trivial measure factor $d\mu[l_a]$ with a simple Lebesgue measure $d^4l_a$. Second, to restrict the integration domain $\Omega$ to the set $\omega$ of the quadruplets $l_a$ given by the distances from a point $P$ in $\mathbb{R}^4$. This is a restriction because a priori, as we stressed in the previous paragraph, the $j_a$'s (and therefore the $l_a$'s) can define a Regge 3d geometry that cannot be embedded in $\mathbb{R}^4$. Therefore $\omega$ is smaller than $\Omega$.

Hence, we define the following integral, which will be the starting point for the rest of our considerations\footnote{With respect to \eqref{cont} we omit the unessential constant $N'(l_{ab})$.}
\be
  I_{+}(l_{ab}):= \int_\omega d^4 l_a
   e^{i\sum_{(ab)}  l_{ab}\delta_{ab}+i\sum_{a} l_a\delta_a}, \label{integ}
\ee
where
\be
\left\{\begin{array}{rl}
\delta_{ab}:=&\pi-\left(\theta_{ab}^c+\theta_{ab}^d\right)\\
\delta_a:=&2\pi-\left(\theta_a^b+\theta_a^c+\theta_a^d\right)
\end{array}\right.
\ee
are the boundary and bulk deficit angles, respectively.

\begin{figure}[h]
\centerline{\includegraphics[width=0.3\textwidth]{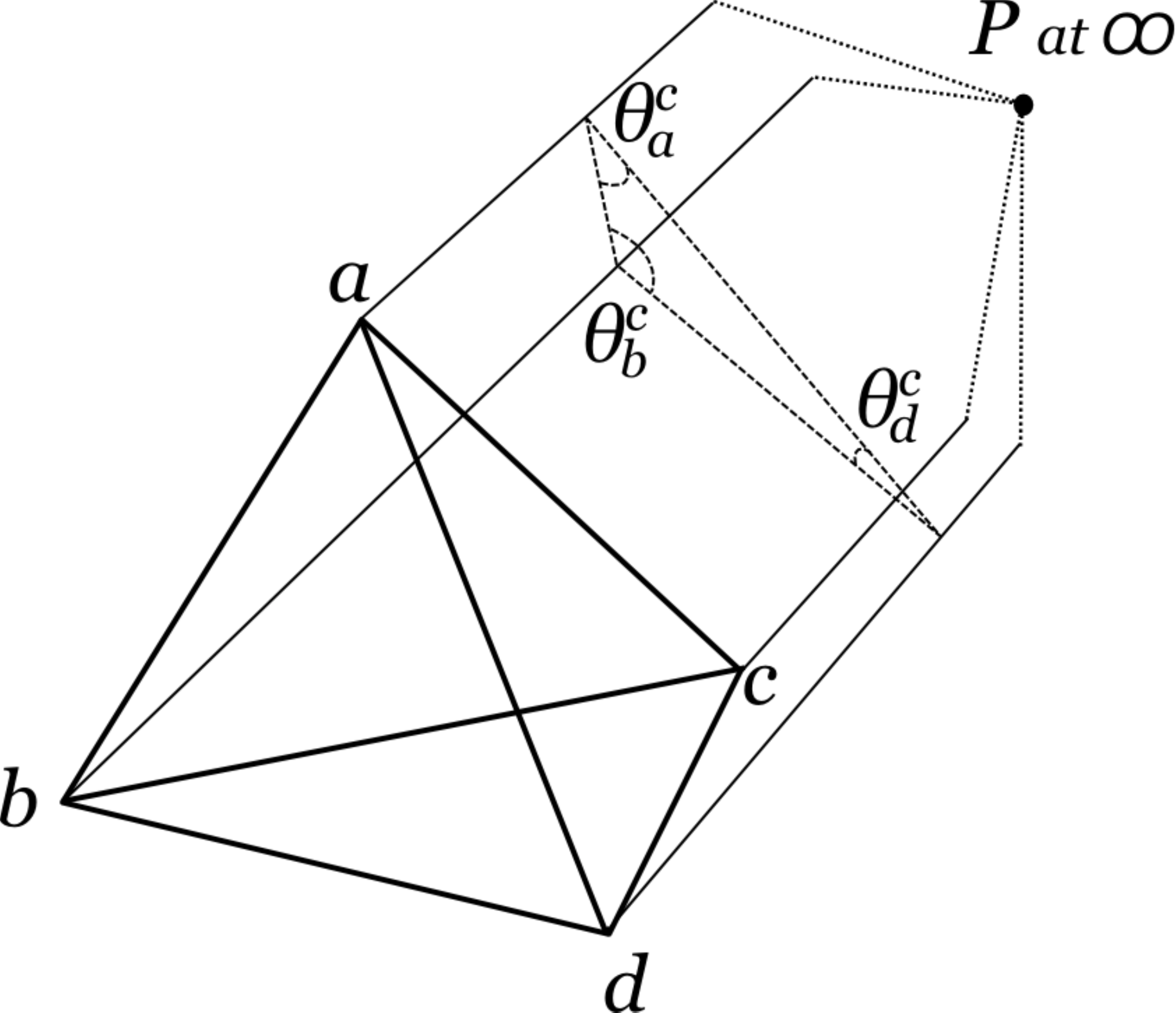}}
\caption{A tetrahedron with an additional fifth vertex at infinity. One can see that the sum of the internal dihedral angles of the same tetrahedron (labelled tetrahedron $c$ above) becomes $\pi$. The same is obviously true for the dihedral angles of the other three tetrahedra. 
\label{asyfig}}
\end{figure}

Switching integration variables from edge lengths $\{l_a\}$, to the 4d coordinates of the point $P$, say $\{x^P_i\}_{i=1,\dots,4}$, and then to spherical coordinates in $\mathbb{R}^4$, that is $\{\rho_P, \alpha^P_l\}_{l=1,\dots,3}$; omitting the label $P$, this yields
\begin{align}
\label{fourDim} 
\left\{ \begin{array}{c c l}
x_1 & =  & \rho\sin\alpha_2\sin\alpha_2\sin\alpha_1 \\
x_2 & =  & \rho\sin\alpha_3\sin\alpha_2\cos\alpha_1 \\
x_3 & =  & \rho\sin\alpha_3\cos\alpha_2 \\
x_4 & = & \rho\cos\alpha_3. \end{array} \right. 
\end{align}
Under this change of variables the integration measure changes in the following way:\footnote{The change in the measure can be easily computed. We start by remarking that 
$$
l_a=\sqrt{(x_1-q_1^a)^2+(x_2-q_2^a)^2+(x_3-q_3^a)^2+(x_4)^2 },
$$
where $q_i^a$ is the $i$-th coordinate in $\mathbb{R}^4$ of the $a$-th vertex of the boundary tetrahedron, and that it is always possible to choose $q\sim l_{ab}$ (of the same order of magnitude as the boundary tetrahedron edges) and $q_4^a=0$. Then, we have:
\begin{align}
 \left\| \frac{\partial l_a}{\partial x_i} \right\|&=\left| \det  \left[\frac{x_i-q_i^a}{l_a}\right] \right|\notag\\
&=\frac{|x_4|}{l_1 l_2 l_3 l_4}\left|\det\Big[(x_1-q^a_1), \cdots , (x_3-q_3^a) , 1\Big]\right|,\notag
\end{align}
and by choosing a linear combination of the columns of the matrix in square brackets, it is easy to see that its determinant is given by a function of the $q_i^a$'s alone. Hence:
$$
d^4l_a=: \rho^{-3}\big[f'(\alpha_l)+g'(\alpha_l,\rho) \big]d^4 x =: \big[ f(\alpha_l)+g(\alpha_l,\rho)\big]d\rho d^3\alpha_l,
$$
where $f(\alpha_l)$ is a function of the angles $\alpha_l$ alone and $g(\alpha_l,\rho)$ is a function that goes to zero as $\rho$ approaches infinity. These functions are regular when $P$ is far from the original tetrahedron (and hence $f(\alpha_l)$ always is). In order to keep the formula cleaner,  we will not deal in the following with the $g(\alpha_l,\rho)$ contributions to the integrals. Nevertheless it is clear that these contributions have a \emph{better} convergence than the $f(\alpha_l)$ ones for large $\rho$; moreover it can also directly be seen that the divergences at $\rho\sim l_{ab}$ are just a coordinate artefact and contribute finitely to the integral.\label{foot}}
\be
d^4l_a = \left\| \frac{\partial l_a}{\partial x_i} \right\| \left\| \frac{\partial x_i}{\partial (\rho,\alpha_l)} \right\|d\rho d^3\alpha_l \approx f(\alpha_l )d\rho d^3\alpha_l. \label{jacob}
\ee
Here, $f(\alpha_l )$ is a function (that we do not need to give explicitly for our purposes) of the angular variables of $S^3$ only, whose typical length scale is given by the size of the boundary tetrahedron.
In these variables, the integral \eqref{integ} takes the form
\begin{align}
I_{+}(l_{ab}) \approx& \int_0^\infty d\rho \int_{S^3} d^3\alpha_l\; f(\alpha_l ) e^{i\sum_{(ab)} l_{ab}\delta_{ab}+i\sum_a l_a\delta_a},\label{regint}
\end{align}
where now all edge lengths and dihedral angles are considered as functions of $\rho$ and $\alpha_l $.
Next, we consider the large $\rho$ asymptotics of the argument of the exponent of \eqref{regint}. The asymptotic form can be best understood by drawing a tetrahedron, adding a fifth vertex $P$ in $\mathbb{R}^4$ and pulling it to infinity (Figure \ref{asyfig}).

We can immediately see that the 4-simplex obtained in this way looks in first approximation as a 4 dimensional prism, with a tetrahedron as its basis. Clearly, the same statement, just transposed into one less dimension, holds also for all the tetrahedra formed by the point $P$ and one of the faces of the original tetrahedron. In such a configuration the sum of the internal dihedral angles belonging to one of these "stretched out" tetrahedra will go to $\pi$ when $P$ approaches infinity:
\be
\text{in $\tau_a$:  }\sum_{b\neq a}  \theta_b^a \xrightarrow{P\rightarrow\infty} \pi.
\ee
Moreover, if $P$ approaches infinity at angles $\alpha_l $ fixed and $\rho\rightarrow\infty$, it is clear that all four edge lengths scale as $l_a=\rho+\text{const.}+\mathcal{O}(\rho^{-1})$. Hence, taking into account all the four tetrahedra, we can easily deduce that our integral takes the following asymptotic form  
\begin{align}
I_{+}(l_{ab}) \approx& \int_0^\infty d\rho \int_{S^3} d^3\alpha_l\; f(\alpha_l ) \exp\Big[i\large(4\pi \rho + F(\alpha_l ) \notag \\
& + G(\alpha_l , \rho)\large)\Big]. \label{reggae}
\end{align}
Here, $F(\alpha_l )$ is solely a function of the angles and $G(\alpha_l ,\rho)$ is the remaining part of the action which goes monotonically to 0 when $\rho$ goes to infinity at fixed $\mathbf{\alpha}$; notice that $G$ does not contain any higher orders of $\rho$ than $\rho^{-1}$. For this expansion to be useful, we must have that $\rho \gg \max\{ l_{ab}\}$, where $l_{ab}$ are the edge length of the fixed (boundary) tetrahedron. In this case, the expansion of the integrand in \eqref{regint} follows directly from the definitions 
of the dihedral angles in our choice of variables.\footnote{Note that the expansion of the dihedral angles related to the boundary edges in \eqref{regint} does not contribute to 
the leading order of the action but to the functions $F(\alpha_l )$ and $G(\alpha_l , \rho)$. Hence, they are clearly included in our expansion as they should.}
The integral \eqref{reggae} is bounded thanks to the oscillations provided by the $4\pi \rho$ term. This fact\footnote{The exact statement that we prove is the following: given any cut-off $C$ on the $\rho$ integration, the norm of the truncated (complex) integral $I_+^C$ is bounded, uniformly in $C$. Symbolically: $$\exists M\;:\;\forall C\quad|I_+^C|<M.$$ } is shown in detail in Appendix \ref{appa}. The result to be kept in mind is that the previous integral is bounded independently on the size of the Regge spike in $\mathbb{R}^4$.\\

Next, we move on to consider the integral $I_{+++-}(l_{ab})$, corresponding to $W_{+++-}$. In this case the integral reads
\begin{align}
I_{+++-}(l_{ab}):=&\int_\omega d^4l_a\;\exp\Big[\dots+i j_4\delta_4+\notag\\
&+\sum_{a\neq4} j_a \left( \theta^4_a- \theta^b_a - \theta^c_a \right)\Big],\label{dihed0}
 \end{align}
where the dots stand for the boundary edge terms, whose details are unessential for our analysis. Manipulating this integral in the same way we did for $I_+$, i.e. going to $\mathbb{R}^4$ spherical variables and taking the $\rho\rightarrow\infty$ limit, it is easy to show that in the action the leading order term in $\rho$ vanishes. Actually, the absence of the leading term in $\rho$ is a peculiarity of the $I_{+++-}$ integral  \eqref{dihed0}; indeed,
it does not happen for the $I_{++--},\; I_{+---}$ or $I_{-}$ cases. In each of them the leading term is $-4\pi \rho, -8\pi \rho$ and $-12\pi \rho$ respectively, and the corresponding integrals are bounded, for the very same reason that $I_+$ is.\footnote{See Appendix \ref{appa}.} This mirrors our geometrical analysis of the saddle points of the $W_{\pm\pm\pm\pm}$ carried out in Section \ref{secsadd}. 

Accordingly \eqref{dihed0} takes the following asymptotic form:
\begin{align}
 I_{+++-}(l_{ab}) \approx &\int_0^\infty d\rho \int_{S^3} d^3\alpha_l\; f(\alpha_l ) \exp\Big[i\large(F_-(\alpha_l ) +\notag\\ &+ G_-(\alpha_l ,\rho)\large)\Big]. \label{reggae2}
\end{align}
Here, as in the $I_+(l_{ab})$ case, the function $F_-(\alpha_l )$ is only a function of the $S^3$ angles
and $G_-(\alpha_l ,\rho)$ is a function that goes monotonically to 0 as $\rho\rightarrow\infty$. By our choice of the measure $d\mu[l_a]\to d^4l_a$, any possible divergence arising in this integral comes from very large $\rho$'s; therefore in looking for divergences we can expand the integrand for $\rho\gg l_{ab}$; in doing this, we get
\begin{align}
f(\alpha_l ) e^{iF_-(\alpha_l )}\Big[1 + iG_-(\alpha_l , \rho)+ \mathcal{O}(\rho^{-2})\Big]. 
 \label{reggdiv}
\end{align}
The above integrand clearly has a linear and a logarithmic divergence when integrated over $\rho$ up to infinity, which are given by its first two terms respectively, while its remainder is convergent.\footnote{Note that the $g(\alpha_l,\rho)$ correction to the measure - see footnote \ref{foot} - can possibly contribute to the logarithmic divergence.} Therefore, 
the integral $I_{+++-}(l_{ab})$ is in fact divergent, as a consequence of the sign flip. 

\section{Summary and Outlook}\label{secconc}

We have pointed out that if we consider the triangulation $^4\tau$ obtained by adding a point $P$ to a tetrahedron $\tau$, as a Regge geometry, the only flat configurations
are those where $P$ is \emph{inside} the boundary of the tetrahedron $\tau$; these flat solutions are all gauge equivalent. 
Any other configuration of $^4\tau$, where the point $P$ is \emph{outside} $\tau$, is \emph{not} flat and carries curvature, depending on the position of the point $P$.

The Ponzano-Regge amplitude of the triangulation defined by $^4\tau$ is well-known to diverge. In the spin representation such divergence scales like $j^3$ in the cut-off spin.  We have analysed the Ponzano-Regge model in its large $j$ limit, where the 6-$j$ symbol is known to behave like the cosine of the Regge action (plus a phase irrelevant here) multiplied by some (crucial) signs.   Since $^4\tau$ is composed by four tetrahedra $\tau_a$, the full amplitude is given by the sum of 16 terms. Each of these can be approximated by a path integral whose action $S_{\pm\pm\pm\pm}$ is - loosely speaking - given by the sum of plus or minus the Regge action for each of the four $\tau_a$'s.   We analysed the classical equation of motion (i.e. the saddle points of the path integral at large $j$'s) for each $S_{\pm\pm\pm\pm}$ and found that there are two main classes of solutions, where the point $P$ lies respectively inside or outside $\tau$. In the first case these are solutions of the $S_{++++}\sim S_{\rm Regge}[^4\tau]$ action, and are  flat and gauge equivalent (in the Regge geometry sense); they form a compact set in the space of the  edge-lengths. In contrast, if $P$ lies outside $\tau$, these are solutions of one of the $(S_{+++-})$-like actions and encode a set of curved, non-gauge-equivalent Regge geometries;  they form a non-compact set in the space of the edge-lengths.  Having found two sets of saddle points for the Ponzano-Regge amplitude of $^4\tau$, we then wanted to explicitly evaluate the path integral of its different terms, to see which ones are at the origin of the divergence and to check if there was any correlation between their divergences and their sets of saddle points.

To be able to evaluate these integrals we had to considerably simplify them. The simplifications consist in trivializing the path integral measure $d\mu[l_a]=\left[\prod_a\frac{2l_a+1}{(12\pi V_a)^{1/2}}\right] d^4 l_a\to d^4 l_a$, as well as in restricting the integration domain to those edge lengths which have an embedding in $\mathbb{R}^4$.

We hence showed that the only divergent path integral is the one related to the $(S_{+++-})$-like actions. Its divergence is linear in the edge-length scale, and not cubic. The discrepancy with the Ponzano-Regge result can be traced to the use of a different measure. In fact each of the $(2l_a+1)$ factors would take into another $l$ factor, while each volume another $l^{-1/2}$ factor.\footnote{Being $\tau$ fixed, and letting the point $P$ go to infinity, one gets a volume that scales like the $l_a$'s (at given $\alpha_l$'s).} The total degree of divergence would hence be $1+4\times1+4\times(-\frac{1}{2})=3$. Unluckily, restoring the original measure entails many complications, which could eventually spoil our result.\footnote{E.g., in the previous naive estimation we did not consider the sector of the integral where the $\alpha_l$ are such that (at least) one of the $V_a$ is zero. If this sector is a source of new divergences depends on the details of the integrand. However, remark that discarding this sector and doing the same kind of estimation for the $W_+$ amplitude, we would find that it does diverge, but with a degree of divergence not larger than 1: the $W_+$ contribution to the total degree of divergence would therefore be subleading.} Despite of this, we think that this calculation can teach us something about the divergences: in fact it shows explicitly, even if in the context of a slightly artificial variation of the Ponzano-Regge model, how the presence of the two signs in the asymptotics can be (one of the) sources of divergences in the large $j$ regime.

These observations are relevant because several spinfoam theories, and in particular  Lorentzian loop quantum gravity in four dimensions \cite{Rovelli:2011eq}, contain the same two terms in the asymptotic expansion of the vertex. This work is in a context of  general effort to understand the possible implications of the presence of these two terms.  For a recent discussion on the physics of the terms, see \cite{Christodoulou:2012sm}. In the language of that paper and \cite{Rovelli:2012yy}, divergences appear to be generated by ``antispacetime" fluctuations of the geometry, or fluctuations of the geometry ``back and forth in time".  

\centerline{---}
\section*{Acknowledgements}

C. Rovelli thanks Daniele Oriti and Sandra Tranquilli for the invitation at the Santa Severa meeting where the initial discussions have (inopportunely) taken place, as well as Bianca Dittrich and Frank Hellmann for many discussion on a previous version of this work. M. L{\aa}ngvik acknowledges grant no. fy2011n45 of the Magnus Ehrnrooth foundation. C. R\"oken was supported by a DFG Research Fellowship.

\appendix

\section{Boundedness of $I_{+}(l_{ab})$ \label{appa}} 
In this appendix we show that the integral  $I_+(l_{ab})$ defined in Equation \eqref{integ} is bounded in an appropriate sense to be specified below.\\

We first define a cut-off version of the integral $I_+(l_{ab})$:
\begin{align}
I_+^C(l_{ab})= & \int_0^C d\rho \int_{S^3} d^3\alpha_l\; f(\alpha_l ) \exp\left[i\sum_{bc}  l_{bc} (\theta^a_{bc}+\theta^d_{bc})\right. \notag \\ 
& \left.+ i\sum_{b} l_{b}(2\pi-(\theta^a_{b} +\theta^c_{b} +\theta^c_{b}))\right].
\end{align}
Then we show that $I_{+}^C(l_{ab})$ is bounded by a number independent of the cut-off $C$. As it will be clear soon, we cannot just say that the whole integral $I_+$ converges because its value is not defined, since it contains an ever-oscillating contribution. To show our claim, we begin with the form \eqref{reggae} of the integral $I^C_{+}(l_{ab})$, and we do the following manipulations to it:
\begin{align}
 I_+^C = & \int_0^C d\rho \int_{S^3} d^3\alpha_l \;f(\alpha_l ) \exp\Big[i(4\pi \rho + F(\alpha_l )+ \notag \\
&+ G(\alpha_l ,\rho))\Big] \\
= &\; 
\int_{S^3}d\mu[\alpha_l] \int_0^C d\rho\;\exp\Big[4\pi i\rho + iG(\alpha_l , \rho)\Big] \label{order} \\
=&\; 
 \int_{S^3}d\mu[\alpha_l]\int_0^C d\rho \;\Big\{\exp({4\pi i\rho})+     \notag      \\
& +\exp({4\pi i\rho})\big[\cos(G(\alpha_l,\rho))-1+i\sin(G(\alpha_l,\rho))\big]\Big\}\\
=&\;
 \mathcal{F}e^{4\pi iC}+\int_{S^3}d\mu[\alpha_l]\int_0^C d\rho\; \Big\{\exp({4\pi i\rho})\times\notag\\
& \times \big[\cos(G(\alpha_l,\rho))-1+i\sin(G(\alpha_l,\rho))\big]\Big\}. \label{a5}
 \end{align}
In equation \eqref{order}, we changed the order of integration and defined $d\mu[\alpha_l ] := f(\alpha_l )\exp[iF(\alpha_l )]d^3\alpha_l$; while in \eqref{a5} we defined $\mathcal{F}:=(4\pi i)^{-1}\int_{S^3}d\mu[\alpha_l]$.\\
Let us now consider the remaining integral in \eqref{a5}. To start, we focus on the integration over $\rho$ at fixed $\alpha_l$'s:
\be
A^C(\alpha_l)=
\int_0^C d\rho\; e^{4\pi i\rho} \big[\cos(G(\alpha_l,\rho))-1+i\sin(G(\alpha_l,\rho))\big].
\ee
We split $A^C$ in real and imaginary part. Each of these is composed of two terms, in which an oscillating function ($\sin(4\pi \rho)$ or $\cos(4\pi \rho)$) multiplies either $\sin(G)$ or $[\cos(G)-1]$. The latter two functions approach monotonically zero for $\rho$ larger than a certain value (which depends on the $\alpha_l$'s), just by the fact that $G$ is an analytic function of $\rho$ whose limit is zero when $\rho$ goes to infinity. By Dirichlet's test (see Appendix \ref{appb} for a reminder) it is immediate to see that each of these terms converge for any $\alpha_l$'s when $C$ goes to infinity. So $A(\alpha_l)=\lim_{C\rightarrow\infty}A^C(\alpha_l)$ are complex numbers of finite norm.\\
Hence:
\begin{align}
\lim_{C\rightarrow\infty}&\Big[I_+^C- \mathcal{F}e^{4\pi iC} \Big] = \lim_{C\rightarrow\infty}\int_{S^3}d^3\mu[\alpha_l]\;A^C(\alpha_l)\notag\\
&=\int_{S^3}d^3\alpha_l\; f(\alpha_l) \exp[iF(\alpha_L)]A(\alpha_l)
\end{align}
which is finite in norm and well-defined, since we are integrating a regular function  over a compact domain.\\
This means that the original integral $I_+(l_a)$, even if not well-defined as a complex number, is still uniformly bounded for any value of the cut-off $C$, i.e. for any size of the spike.
 
\section{Dirichlet's Test \label{appb}}

Dirichlet's test for the convergence of integrals goes as follows.\\
 If two functions $f(x)$ and $g(x)$ are such that 
\be
\left|\int_a^y dx\; f(x)\right| <  M, \;\forall y>a 
\ee
for a certain real fixed $M$, and
\be
 g(x)\xrightarrow{x\rightarrow\infty}0 \;\text{monotonically for } x>a,
\ee 
then the integral of their product converges:
\be
\left|\int_a^\infty dx \;f(x)g(x)\right|<\infty.
\ee


\vfill

\begin{thebibliography}{10}
\section*{References}

\bibitem{Rovelli:2011eq}
C.~Rovelli, ``{Zakopane lectures on loop gravity},''
\href{http://arxiv.org/abs/1102.3660}{{\tt arXiv:1102.3660}}.

\bibitem{Barrett:2009mw}
J.~W. Barrett, R.~Dowdall, W.~J. Fairbairn, F.~Hellmann, and R.~Pereira,
  ``{Lorentzian spin foam amplitudes: Graphical calculus and asymptotics},''
  \href{http://dx.doi.org/10.1088/0264-9381/27/16/165009}{{\em
  Class.Quant.Grav.} {\bf 27} (2010)  165009},
\href{http://arxiv.org/abs/0907.2440}{{\tt arXiv:0907.2440}}.

\bibitem{Freidel:2004vi}
L.~Freidel and D.~Louapre, ``Ponzano--Regge model revisited. I: Gauge fixing,
  observables and interacting spinning particles,'' {\em Class. Quantum Grav.}
  {\bf 21} (2004)  5685.

\bibitem{Baratin:2011tg}
A.~Baratin, F.~Girelli, and D.~Oriti, ``{Diffeomorphisms in group field
  theories},'' \href{http://arxiv.org/abs/1101.0590}{{\tt arXiv:1101.0590}}.

\bibitem{Bahr:2009mc}
B.~Bahr and B.~Dittrich, ``{Breaking and restoring of diffeomorphism symmetry
  in discrete gravity},''
\href{http://arxiv.org/abs/0909.5688}{{\tt arXiv:0909.5688}}.

\bibitem{Fairbairn:2010cp}
W.~J. Fairbairn and C.~Meusburger, ``{Quantum deformation of two
  four-dimensional spin foam models},''
\href{http://arxiv.org/abs/1012.4784}{{\tt arXiv:1012.4784}}.

\bibitem{Han:2011vn}
M.~Han, ``Cosmological Constant in LQG Vertex Amplitude,''
  \href{http://arxiv.org/abs/1105.2212}{{\tt arXiv:1105.2212}}.


\bibitem{Rovelli:2011mf}
C.~Rovelli, ``{On the structure of a background independent quantum theory:
  Hamilton function, transition amplitudes, classical limit and continuous
  limit},''
\href{http://arxiv.org/abs/1108.0832}{{\tt arXiv:1108.0832}}.

\newpage

\bibitem{Freidel:2002dw}
L.~Freidel and D.~Louapre, ``{Diffeomorphisms and spin foam models},''
  \href{http://dx.doi.org/10.1016/S0550-3213(03)00306-7}{{\em Nucl. Phys.} {\bf
  B662} (2003)  279--298},
\href{http://arxiv.org/abs/gr-qc/0212001}{{\tt arXiv:gr-qc/0212001}}.

\bibitem{Christodoulou:2012sm}
M.~Christodoulou, A.~Riello, and C.~Rovelli, ``{How to detect an
  anti-spacetime},''
\href{http://arxiv.org/abs/1206.3903}{{\tt arXiv:1206.3903}}.

\bibitem{Ponzano:1968uq}
G.~Ponzano and T.~Regge, ``Semiclassical limit of Racah coefficients,'' in {\em
  Spectroscopy and group theoretical methods in Physics}, F.~Bloch, ed.
\newblock North-Holland, Amsterdam, 1968.

\bibitem{Morse:1991te}
P.~A. Morse, ``{Approximate diffeomorphism invariance in near flat simplicial
  geometries},''. UCSBTH-91-46.

\bibitem{Rovelli:2012yy}
C.~Rovelli and E.~Wilson-Ewing, ``{Discrete Symmetries in Covariant LQG},''
\href{http://arxiv.org/abs/1205.0733}{{\tt arXiv:1205.0733}}.

\bibitem{Engle:2011ps}
J.~Engle, ``{The Plebanski sectors of the EPRL vertex},'' {\em
  Class.Quant.Grav.} {\bf 28} (2011)  225003,
\href{http://arxiv.org/abs/1107.0709}{{\tt arXiv:1107.0709}}.

\bibitem{Engle:2011un}
J.~Engle, ``{A proposed proper EPRL vertex amplitude},''
\href{http://arxiv.org/abs/1111.2865}{{\tt arXiv:1111.2865}}.

\bibitem{Engle:2012yg}
J.~Engle, ``{A spin-foam vertex amplitude with the correct semiclassical
  limit},''
\href{http://arxiv.org/abs/1201.2187}{{\tt arXiv:1201.2187}}.

\bibitem{Barrett:2008wh}
J.~W. Barrett and I.~Naish-Guzman, ``{The Ponzano-Regge model},''
  \href{http://dx.doi.org/10.1088/0264-9381/26/15/155014}{{\em
  Class.Quant.Grav.} {\bf 26} (2009)  155014},
\href{http://arxiv.org/abs/0803.3319}{{\tt arXiv:0803.3319}}.

\bibitem{CM}
D.~S. Mitrinovic and J. Pecaric and V. Volenec,  ``{Recent Advances in
Geometric Inequalities},'' Mathematics and its Applications.


\end{thebibliography}
\end{document}